\documentclass[%
 ,twocolumn
, showpacs
,amssymb, amsmath,nobibnotes, aps, prl]{revtex4}

\usepackage{graphicx}
\usepackage{dcolumn}
\usepackage{bm}

\begin{document}
\title{
Disorder effect on the localization/delocalization in incommensurate potential 
}
%
\author{Masaru Onoda$^{1,2}$}
\email{m.onoda@aist.go.jp}
\author{Naoto Nagaosa$^{1,2,3}$}
\email{nagaosa@appi.t.u-tokyo.ac.jp}
\affiliation{
$^1$Correlated Electron Research Center (CERC),
National Institute of Advanced Industrial Science and Technology (AIST),
Tsukuba Central 4, Tsukuba 305-8562, Japan\\
$^2$CREST, Japan Science and Technology Corporation (JST), Saitama, 332-0012, Japan\\
$^3$Department of Applied Physics, University of Tokyo, 
Bunkyo-ku, Tokyo 113-8656, Japan
}
%
%
\date{\today}
%
\begin{abstract}
The interplay between incommensurate (IC) and random potentials
is studied in a two-dimensional symplectic model with the focus on
localization/delocalization problem. With the IC potential only, 
there appear wavefunctions localized along the IC wavevector while extended 
perpendicular to it. Once the disorder potential is introduced, 
these turn into two-dimensional anisotropic metallic states beyond the scale 
of the elastic mean free path, and eventually becomes localized in both 
directions at a critical strength of the disorder.
Implications of these results to the experimental observation 
of the IC-induced localization is discussed.
\end{abstract}
\pacs{
72.15.Rn, 
73.20.Fz, 
73.20.Jc, 
64.70.Rh, 
}
\maketitle

The localization/delocalization of the electronic wavefunctions is one of 
the most fundamental problems in condensed matter physics. In the presence 
of the disorder, the interference among the scattered waves leads to the 
quantum correction to the Boltzmann transport theory and eventually to the 
Anderson localization \cite{Lee}. This is most 
prominent in one-dimension, where all the states are localized for any finite 
strength of the disorder potential. There is another type of localization, 
which is induced by the incommensurate (IC) potential as first discussed 
by Aubry-Andre\cite{AuA,Sokoloff}. In this case, even in one-dimension, 
there is a critical strength of the IC potential 
at which the localization/delocalization transition 
occurs for all the energies simultaneously. 
This suggests that these two types of localization are different 
in nature even though the quantum interference is the key issue 
in both cases.  
 
  It is rather easy to find the disordered system showing the 
Anderson localization, while there has been no 
experimental report on the IC-induced localization.
Charge (spin) density wave system \cite{RMP60_001129_88,RMP66_000025_94}, 
helical magnets \cite{Tanaka}, 
quasi-crystal \cite{PRL50_001870_83,PRB35_001020_87}, 
and two-dimensional electrons under magnetic field (Hofstadter problem)
\cite{PRB14_002239_76}
are the representative candidates for it. In the realistic systems, however, 
there is always some degree of disorder and it is crucial to see the 
effect of the disorder potential in addition to the IC
potential, which we address in this letter.

When one considers the IC potential with one 
wavevector $\bm{q}_{\mathrm{IC}}$, which we take 
along $x$-axis, the translational symmetry along the two directions $y$, 
$z$ perpendicular to $\bm{q}_{\mathrm{IC}}$ 
remains intact, and the momenta $k_y$, $k_z$ are 
well-defined. Therefore the localization occurs only 
along $x$-direction and the wavefunctions are extended along $y$, $z$ directions. 
With the disorder, one possible scenario is that 
the two-dimensional metallic state induced by IC potential is 
localized by infinitesimal disorder due to the Anderson localization.
However, this separation into one and two dimensions breaks down 
when the disorder is introduced. 
Namely the scatterings by random potential reduce the spatial anisotropy. 
Then a keen question is the localization/delocalization in the presence 
of both IC and disorder potentials. 
Therefore, the second scenario is that the three-dimensional metal 
is recovered by the disorder potential.  
Similar problem arises for the two-dimensional symplectic models
which remain metallic for weak disorder strength. 

In this paper, we consider a two-dimensional model with spin-orbit 
interaction. This model belongs to the symplectic universality class and 
shows a metal-insulator transition at a finite critical strength of 
disorder without the IC potential, similarly to the three-dimensional case. 
With the IC potential but without the disorder, there appear
localized states along $\bm{q}_{\mathrm{IC}}$-direction
at least for some energy region. 
The Hamiltonian of this model is given by
\begin{eqnarray}
H 
&=&
H_{0}+V 
\\
H_{0}
&=&
\sum_{\bm{r},\bm{r}'}
c^{\dagger}_{\bm{r}}t_{\bm{r}\bm{r}'}c_{\bm{r}'},
\\
t_{\bm{r}\bm{r}'}
&=& \left\{
\begin{array}{rl}
\sqrt{1-S^2}t\pm iSt\sigma_{y}, & \bm{r} = \bm{r}' \pm a\bm{e}_{x}\\
\sqrt{1-S^2}t\mp iSt\sigma_{x}, & \bm{r} = \bm{r}' \pm a\bm{e}_{y}\\
\end{array}
\right. ,
\label{eq:trr}
\\
V &=&
\sum_{\bm{r}} 
c^{\dagger}_{\bm{r}}\left[\frac{U}{2}\sin(\bm{q}_{\mathrm{IC}} \cdot \bm{r})
+w_{\bm{r}}\right]c_{\bm{r}},
\end{eqnarray}
where 
$c^\dagger_{\bm{r}}= ( c^\dagger_{\bm{r} \uparrow},
c^\dagger_{\bm{r} \downarrow})$
is the two-component creation operator, while $c_{\bm{r}}$
is its hermitian conjugate annihilation operator.
The term with $S$ in the transfer integral in Eq.~(\ref{eq:trr}) represents
the spin-orbit interaction, $U$ is the IC potential, and
$ w_{\bm{r}}$ is the random disorder potential.  
The case of $U=0$ has been studied by Ando \cite{PRB40_005325_89}.
When the Fermi energy is near the band top or bottom,
the non-perturbative part $H_{0}$ reduces to the Rashba model.
In the case of $S\ll 1$, the effective mass $m^{*}$ 
and the Rashba coupling $\alpha$ are given by $m^{*} \sim 1/(2ta^{2})$ 
and $\alpha \sim 2Sta$, respectively.
We shall take $S=0.5$ as in Ref.~\cite{PRB40_005325_89}
and focus on the energy level around the band center
in the rest of this paper. 
The IC wavevector $\bm{q}_{\mathrm{IC}}=(2\pi/\lambda_{\mathrm{IC}})\bm{e}_{x}$
is taken along the $x$-direction with the wavelength $\lambda_{\mathrm{IC}}$.
The on-site potential $w_{\bm{r}}$ is randomly distributed in
the range $[-W/2, W/2]$.
In the following we shall consider the IC potential
with $U=4.0t$ and $\lambda_{\mathrm{IC}} = \sqrt{5}a$.
The energy unit (hopping parameter) $t$ and the length unit 
(lattice constant) $a$ are set to be 1 hereafter.

\begin{figure}[t]
  \includegraphics[scale=0.35]{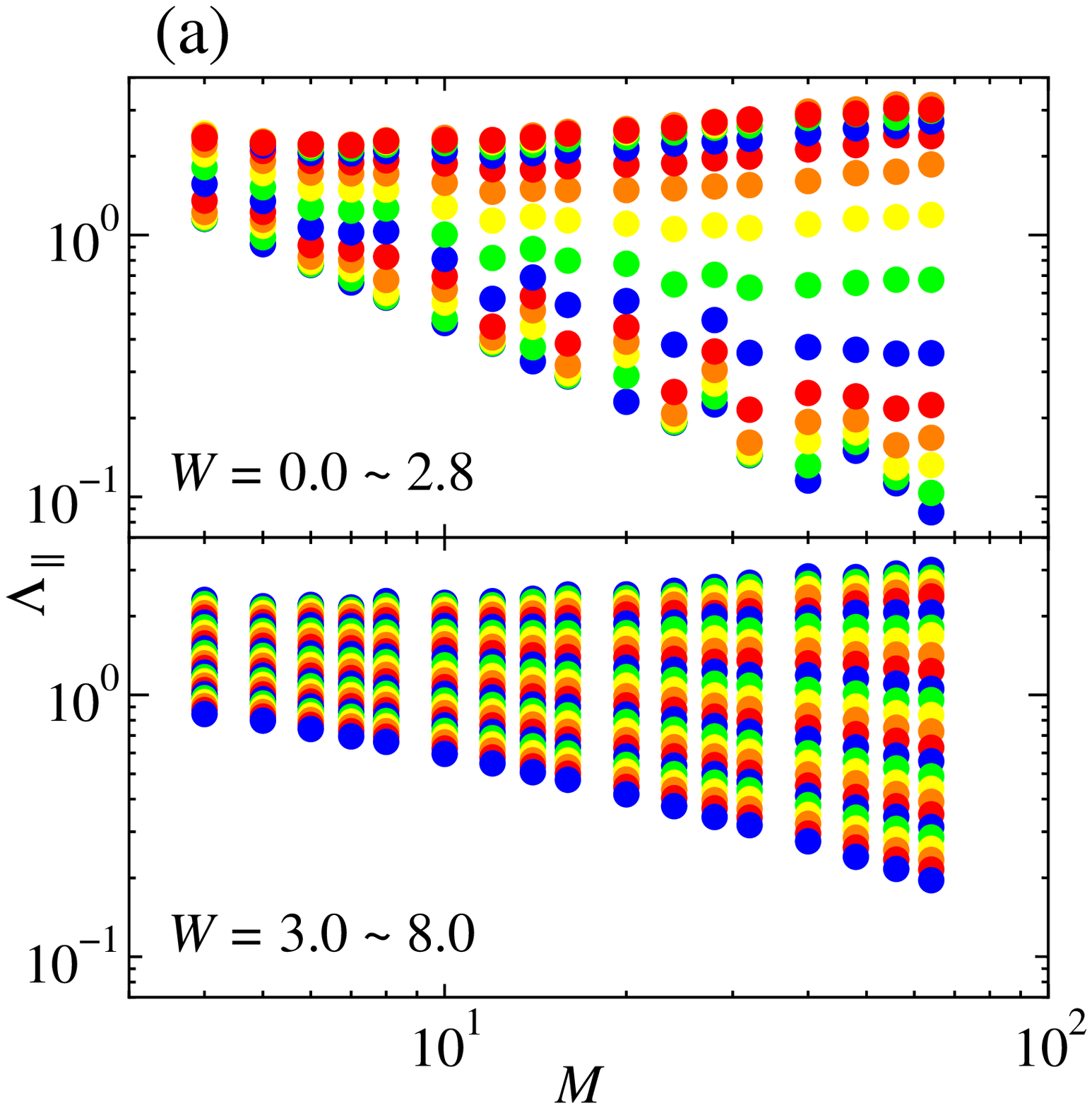}
  \includegraphics[scale=0.35]{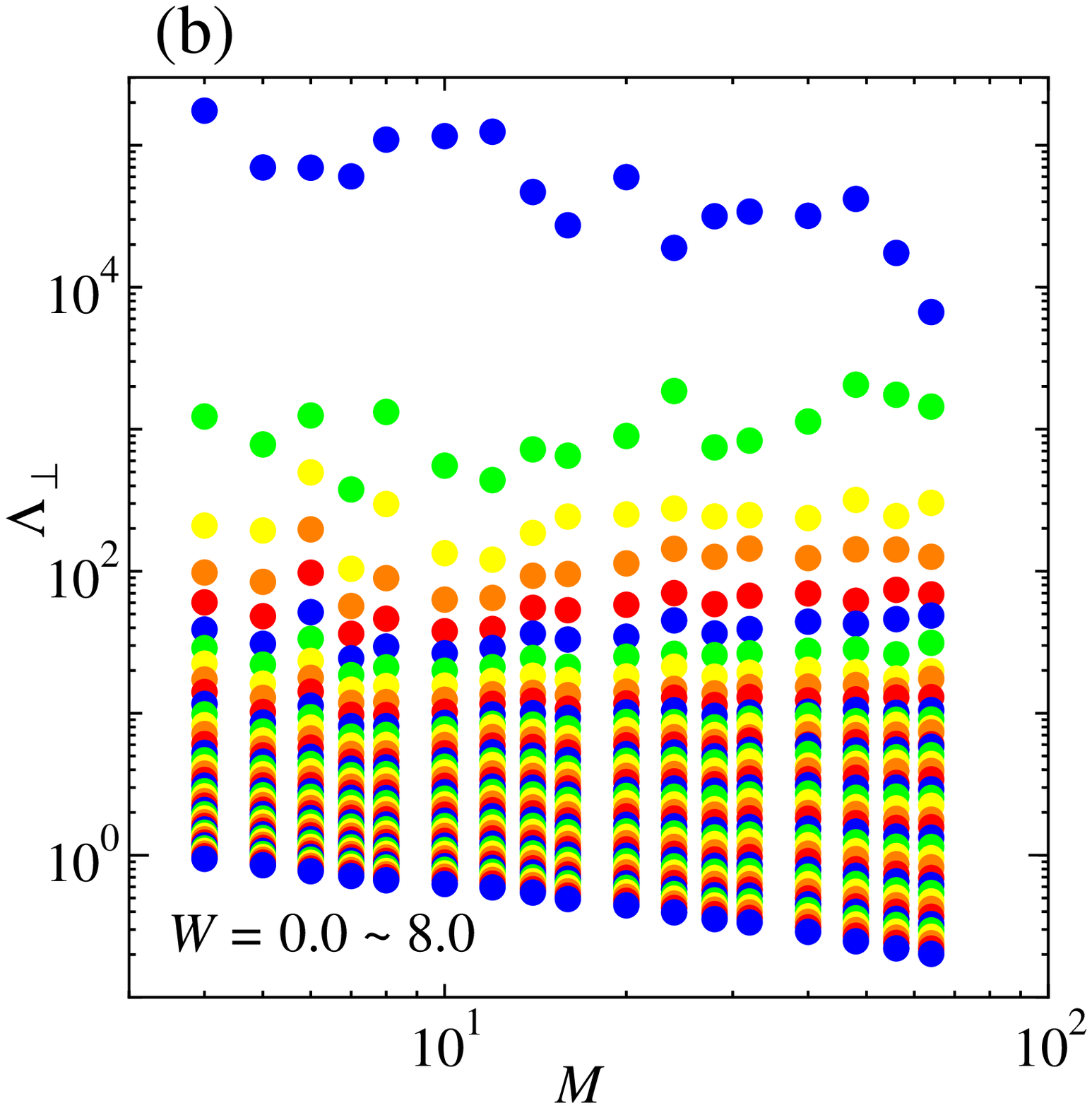}
  \caption{
The renormalized localization (correlation) length,
$\Lambda_\alpha(M,W) = \lambda_{\alpha M}(W)/M$, 
for (a) parallel ($\alpha=\parallel$) and 
(b) perpendicular ($\alpha=\perp$) 
direction to $\bm{q}_{\mathrm{IC}}$,
at the band center in the system with 
incommensurate potential of $U = 4.0$ and
$\lambda_{\mathrm{IC}} = \sqrt{5}$.
}
\label{fig:raw}
\end{figure}
We have calculated the localization length $\lambda_{\alpha M}(W)$
of the above model in the geometry of quasi-one dimensional tube
with  $M$-site circumference 
in terms of the transfer matrix method~\cite{ZPB53_0000001_83}.
We consider the cases in which
the direction of tube is parallel ($\alpha = \parallel$) or
perpendicular ($\alpha=\perp$) to the IC wavevector $\bm{q}_{\mathrm{IC}}$.
The system without IC potential is also analyzed for comparison.
Figures~\ref{fig:raw}(a) and  \ref{fig:raw}(b) show the renormalized localization length 
$\Lambda_\alpha(M,W) = \lambda_{\alpha M}(W)/M$  
as a function of the circumference $M$ 
for several values of the disorder strength $W$.
The length of a tube is taken to be $N=4\times10^{5}$ sites.
We have calculated $\Lambda_\alpha(M,W)$ with $M=$
4, 5, 6, 7, 8, 10, 12, 14, 16, 20, 24, 28, 32, 40, 48, 56, and 64, 
and $W$ is changed from 0.0 to 8.0 with the interval of 0.2.
From the variation of the data for $1/\lambda_{\alpha M}$, we have estimated 
its error, which is at most a few percent at $\lambda_{\alpha M}<100$.
This means that the error for $\Lambda_{\parallel}$ in Fig.~\ref{fig:raw}(a) 
is at most a few percent, while it is comparable to the average
for the data $\Lambda_{\perp} > 2 \times 10^1$ in Fig.~\ref{fig:raw}(b),
as suggested also by their scattered behavior.
We present these data for $\Lambda_{\perp}>2 \times 10^1$
to show only the systematic tendency. 
In Fig.~\ref{fig:raw}(b), $\Lambda_\perp$ is monotonously 
decreasing with increasing $W$. On the other hand,
$\Lambda_\parallel$ is nonmonotonic as shown in Fig.~\ref{fig:raw}(a);
in the upper panel of Fig.~\ref{fig:raw}(a),
the disorder potential $W$ increases from bottom to the top,
while it is reversed in the lower panel.
The lowest curve in the upper panel of Fig.~\ref{fig:raw}(a) 
corresponds to $W=0$, where only the IC potential
is at work and the wavefunction is strongly localized
along $\bm{q}_{\mathrm{IC}}$, while it is extended perpendicular to
it since $k_y$ is a good quantum number. 
( The top curve in Fig.~\ref{fig:raw}(b)
shows that $\lambda_{\perp M}$ reaches the sample size $N=4 \times 10^5$.) 

\begin{figure}[b]
  \includegraphics[scale=0.35]{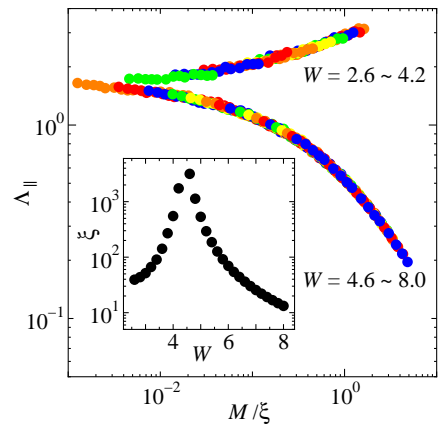}
  \caption{
Renormalized localization (correlation) length,
$\Lambda_{\parallel}(M,W) = \lambda_{\parallel M}(W)/M$,
parallel to $\bm{q}_{\mathrm{IC}}$ as a function of $M/\xi(W)$ 
where $\xi(W)$ is the localization (correlation) length
in the thermodynamic limit.
(Inset) the localization (correlation) length $\xi(W)$
as a function of $W$.
}
\label{fig:scaling}
\end{figure}

\begin{figure*}[hbt]
  \includegraphics[scale=0.35]{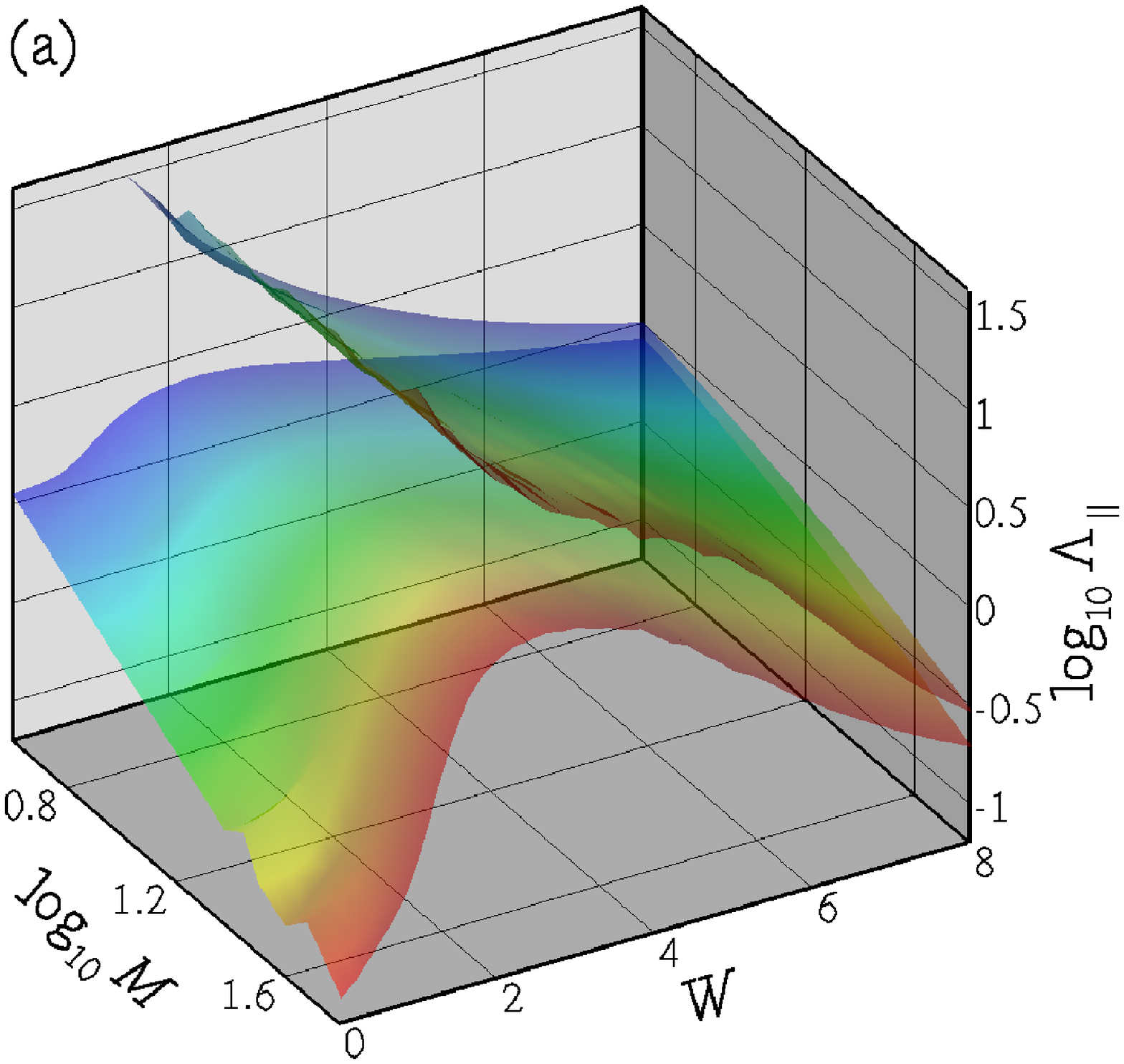}
  \includegraphics[scale=0.35]{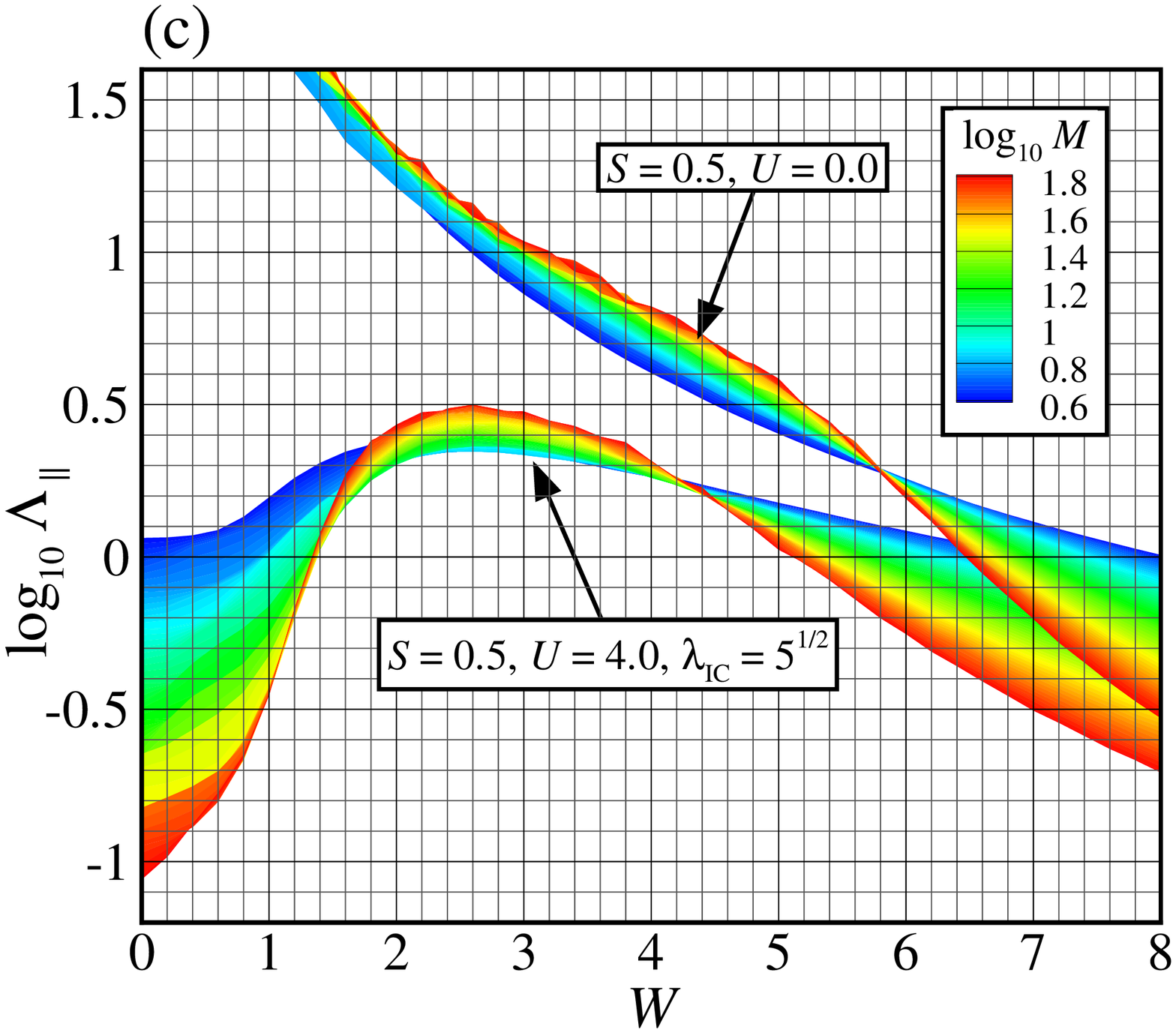}
  \includegraphics[scale=0.35]{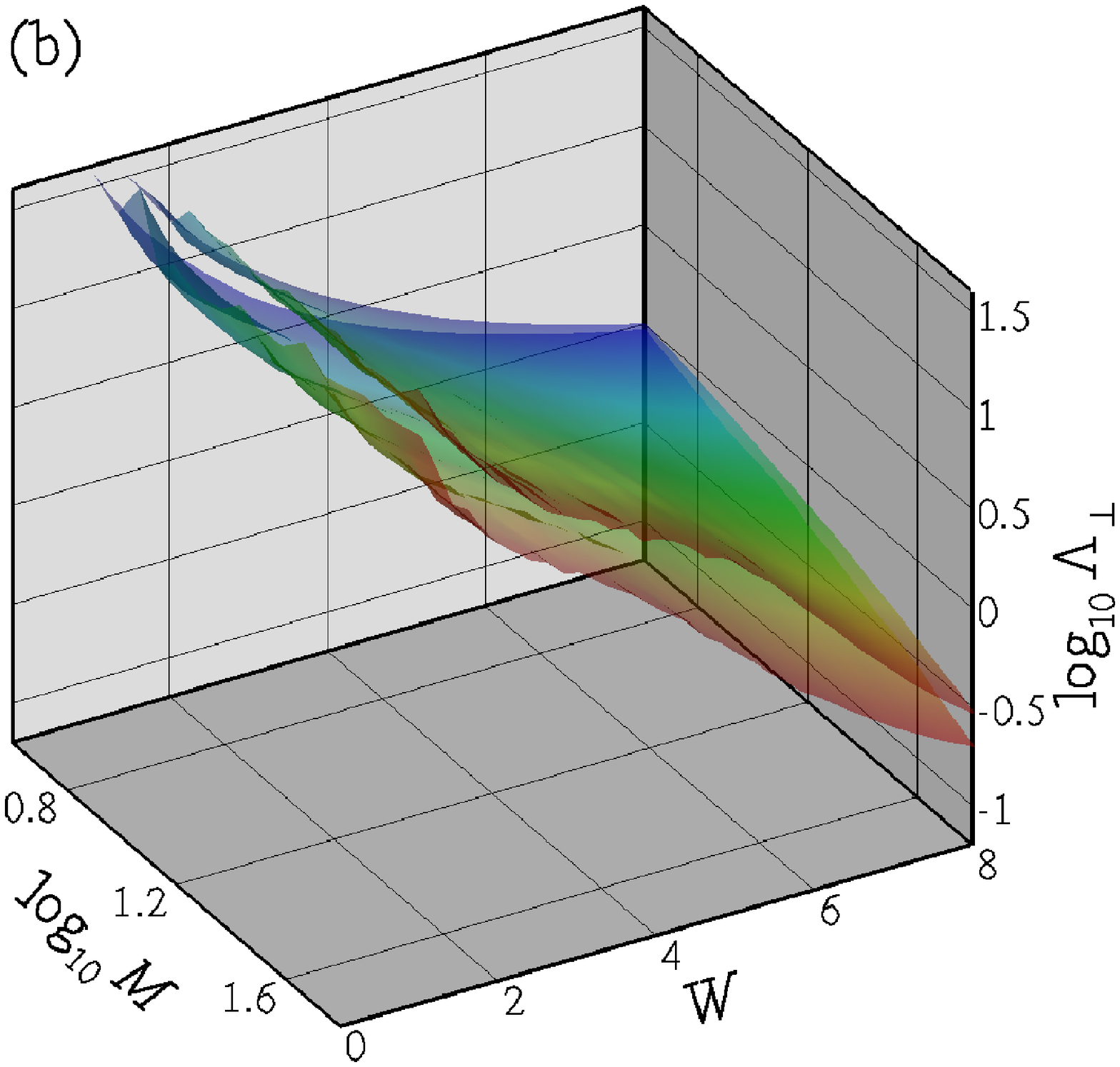}
  \includegraphics[scale=0.35]{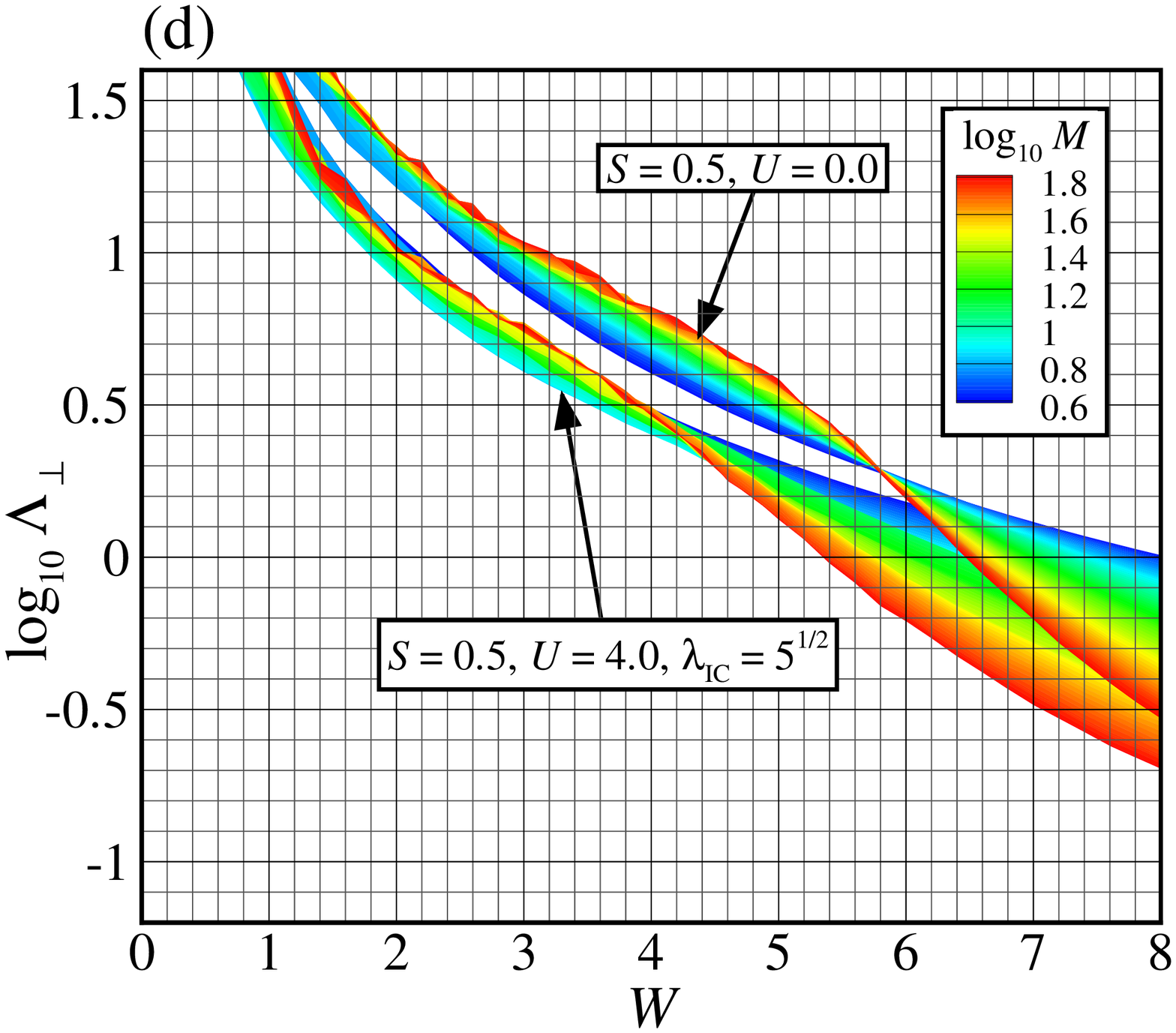}
  \caption{
Birds-eye view of
renormalized localization (correlation) length,
$\Lambda_{\alpha}(M,W) = \lambda_{\alpha M}(W)/M$, 
for (a) parallel ($\alpha=\parallel$) and 
(b) perpendicular ($\alpha=\perp$) 
direction to $\bm{q}_{\mathrm{IC}}$,
at the band center as a function of $W$ and $M$.
In each of (a) and (b),the upper and lower planes
represents the cases without and with the IC potential, respectively.
The projections of (a) and (b) to the $W$-$\Lambda$ plane
are shown in (c) and (d), respectively.
}
\label{fig:bird}
\end{figure*}
\printfigures

With finite $W<2.8$ , $\Lambda_\parallel$
shows a nonmonotonous behavior as a function of $M$;
it decreases for $M < M_c(W)$ and turns to 
increase for $M>M_c(W)$. This means that there 
appears a crossover length scale $M_c(W)$,
which is a decreasing function of $W$, 
from localized to extended states along 
$\bm{q}_{\mathrm{IC}}$. In this range of $W$ and $M$ we have studied,
it is clear that the single-parameter scaling does not
hold, and it is not clear if $M_c(W)$ diverges 
as $W \to W_c >0$ or $W \to 0$. However, it is 
convincing that the originally localized state along 
$\bm{q}_{\mathrm{IC}}$-direction turns to be extended with
the disorder $W$. 
A tentative estimate of $M_c(W)$ from the upper panel
of Fig.~\ref{fig:raw}(a) is $M_c(W=1.2)  \sim 40$ 
(7-th green curve from the bottom ), 
$M_c(W=1.6) \sim 20$
(9-th orange curve from the bottom),
$M_c(W=2.0) \sim 10$
(11-th blue curve from the bottom).
These values are roughly consistent with
the relation $M_c(W) \propto W^{-2}$,
which suggests that $M_c(W)$ scales 
with the elastic mean free path due to the 
disorder, beyond which the incommensurability is 
lost and the system behaves as an anisotropic two-dimensional system. 
For weak disorder, 
the elastic mean free path $\ell(W)$ is roughly estimated as 
$\ell(W)\sim 48a(t/W)^2$.
The above formula is derived from the estimations of
the mean free time $\tau$ as
$\tau^{-1}\sim \pi\langle w^{2}_{\bm{r}}\rangle N_{F}$,
the density of states per unite cell
$N_{F}\sim 1/(2\pi t)$,
the averaged disorder strength 
$\langle w^{2}_{\bm{r}}\rangle \sim W^2/12$ for 
the rectangular distribution,
and the Fermi velocity $v_{F}\sim 2ta$ around the band center.
This estimation semi-quantitatively supports
the identification, $M_c(W)\sim\ell(W)$,
since $\ell(W=1.2)\sim 33$, $\ell(W=1.6)\sim 19$ 
and $\ell(W=2.0)\sim 12$, for this regime of disorder.

In the lower panel of Fig.~\ref{fig:raw}(a), on the other hand, $W$
increases from the top ($W=3.0$) to the bottom ($W=8.0$). 
At $W=3.0$, $M_c(W)$ is almost the lattice constant 
and $\Lambda_\parallel$ shows the monotonous behavior 
as a function of $M$ for $W>3.0$.
This is similar to $\Lambda_\perp$ in Fig.~\ref{fig:raw}(b), and 
is the canonical behavior of the metal-insulator transition. 
Therefore we show in Fig.~\ref{fig:scaling} its single-parameter
scaling analysis \cite{PRL42_000673_79}
from the lower panel of Fig.~\ref{fig:raw}(a). 
The degeneracy of the data
indicating the scaling relation
\begin{equation}
\Lambda_\parallel = f_\parallel\left(\frac{M}{\xi}\right)
\end{equation}
is clearly obtained and the resultant 
localization/correlation length $\xi(W)$ is plotted in the 
inset of Fig.~\ref{fig:scaling}.
(In the upper branch, only the data for $M\geqq 8$ are used.)

The divergence of $\xi(W)$ toward $W_{c}= 4.3\pm 0.1$ 
does not contradict with the relation
$\xi(W) \sim | W- W_c|^{-\nu}$ with the 
known critical exponent $\nu = 2.7\pm 0.1$, although
we did not try the accurate estimate of $\nu$.
Therefore we believe that the present metal-insulator transition 
belongs to the universality class of  
two-dimensional symplectic one
\cite{PRB40_005325_89,PhysA172_302_91,PRL89_256601_02}.

In order to obtain a global picture, we plot in 
Figures~\ref{fig:bird}(a) and \ref{fig:bird}(b) 
the birds-eye view of the renormalized
localization length $\Lambda_\alpha(M,W) = \lambda_{\alpha M}(W)/M$
as a function of the circumference $M$ and the disorder strength $W$
for (a) $\alpha=\parallel$ and (b) $\alpha=\perp$.
In Figs.~\ref{fig:bird}(c) and \ref{fig:bird}(d), 
we also put the projections of Figs.~\ref{fig:bird}(a) and \ref{fig:bird}(b)
to the $W$-$\Lambda_\alpha$ plane, respectively.
In each panel, two curved surfaces are shown.
The upper (lower) surfaces are for the case without (with) IC  potential.
It should be noted that
the color contour represents the value of $M$ in logarithmic scale,
different from the colors in Figs.~\ref{fig:raw} and \ref{fig:scaling}, 
in order to make the connection to the 
projected figures on the righthand side.  
From the $\Lambda_{\parallel}$ with $U=4.0$ 
in Figs.~\ref{fig:bird}(a) and \ref{fig:bird}(c), 
it is clearly seen that the state is localized,
i.e., the red part is lower than the blue one,
at both small $W \cong 0$ and large $W >W_c = 4.3\pm 0.1$.
The degeneracy occurs at $W_c$ indicates the 
metal-insulator transition as analyzed in 
Figs.~\ref{fig:raw} and \ref{fig:scaling},
while one can not see any degeneracy for smaller $W$. 
Therefore, the reversal of the $M$-dependence around 
$W\cong 1.5$ is attributed to the crossover at 
$M_c(W)$ as discussed above.
The behavior of $\Lambda_\perp$ 
in Figs.~\ref{fig:bird}(b) and \ref{fig:bird}(d), 
on the other hand, is similar to the case of $U=0$.
Note that the critical $W_c$ is common for both
$\Lambda_{\parallel}$ and $\Lambda_{\perp}$,
which is reduced from $W_c= 5.8\pm 0.1$ to $W_c= 4.3\pm 0.1$
by the IC potential $U=4.0$.

Now we discuss the relevance of these results to 
the experimental observation of the localization 
due to IC potential. 
Except the one-dimensionally confined systems such as 
quantum wire, the electronic wavefunctions are extended 
in the perpendicular directions to the IC wavevector. 
This is almost always the case as in charge (spin) density 
wave, and helical magnets. 
In this case, we expect that the elastic scattering by impurities 
washes out the IC potential in both two and three dimensions
beyond the length scale of the elastic mean free path $\ell(W)$.
First let us ask the following question: `` Can one observe the 
temperature dependence of the resistivities
characteristic to the IC-induced localization ? ``
To answer to this question, one should remember 
that the temperature dependence of the 
resistivity is translated into
that of the inelastic mean free path $L_{\rm in}(T)$,
which is replaced by the sample size $L$. 
The quantum correction due to the interference occurs
for the length scale $L > \ell(W)$.
Namely, $\ell(W)$ is the shortest length scale
for the localization problem, and the Boltzmann
transport theory applies for $L<\ell(W)$.
Therefore, if the crossover length scale $M_c(W)$
is identical to $\ell(W)$, as suggested by the 
analysis above, there is no chance to observe 
the IC-induced localization from the temperature dependence.
Most probably this is the reason why there are no
experimental observation of the IC-induced localization,
i.e., it is very fragile against the disorder and 
is replaced by the usual Anderson localization.
However, the incommensurability gives a strong anisotropy of
the metallic conduction between parallel and perpendicular
directions 
to the wavevector $\bm{q}_{\mathrm{IC}}$ within the 
Boltzmann transport regime,
which is the remnant of the IC-induced localization.
Therefore, for example, one should look for the dramatic change of the 
Resistivity anisotropy associated with the commensurate-incommensurate
transition.

We acknowledge H.~Katsura and S.~Tanaka for the
useful discussion.
This work is financially supported by NAREGI Grant,
Grant-in-Aids from the Ministry of Education,
Culture, Sports, Science and Technology of Japan.

\end{document}